# On Calculation of Lattice Energy in Spatially Confined Domains

Yevgen Bilotsky

Department of Material Science and Engineering, Aalto University Foundation School of Chemical Technology,
P.O. Box 16200, FIN-00076 AALTO, Finland

yevgen.bilotsky@aalto.fi

*Abstract*-Evaluation of internal energy and the inter-atomic or ionic interactions in a crystal lattice usually requires precise calculation of lattice sums. This in the case of small nano-particles (as space-limited domains) presents several challenges, as conventional methods are usually valid only for infinite lattices, tailored for a specific potential. In this work, a new method has been developed for calculation of atomic interactions based on the radial density function with the geometric probability approach, extended to arbitrary fixed lattices and potentials in a nano-particle.

The derived radial density function (RDF) combines terms for a uniform particles distribution, for non-uniform spherical symmetry and the last one for an additional, angle-dependent term. The second term originates from Walfisz-like formula for lattice sums. The RDF with these three terms is explicitly integrated for spherical lattice domains resulting in the internal energy of the system with a prescribed interaction potential. The application of the method was demonstrated for Wigner model of electrons lattice interacting with compensating positive jelly in finite lattice sphere, which interacting energy between lattice and jelly was evaluated. The excess of this energy caused by space-limitation of the lattice was explicitly expressed in the terms of absolutely convergent lattice sums.

I. INTRODUCTION

Many properties of nano-size scale materials are rather distinct from those of bulk materials. However, in the calculation of thermodynamic properties such as internal energy of nano-particles, the same approach as for bulk (infinite) systems is used, where only the extra surface energy contribution is being added [1]. The alternate approach for calculation of long-range Coulomb interaction in a large finite crystal in systems, consisting of periodically repeated replicas of identical unit cells, has been proposed in [2] (together with useful citations on that subject). In that article, by using some additional algebra, the final result expressed there as a six-dimensional integral can be rearranged into a much simpler form given by a 3D integral. Mathematical difficulties, arising from conditional convergence of the Coulomb series in finite crystals, have been analysed in [3]. It is worth noting that using potentials other than Coulomb potential requires essential modification for calculation of lattice sums in conventional approach.

In this work, we suggest a new analytical approach for the calculation of the energy of finite (space-limited) nano-particles with an arbitrary type of crystalline lattice. The interaction energy between two points (atoms, charges etc.), used here, is based on the analytical central interaction $U(|\vec{r}_1 - \vec{r}_2|)$. In this approach, we directly calculate the numbers of pairs of interacting points as the function of distances between these points in the domain. In this case, addition of the surface energy term is not required, inasmuch as the surface contribution is automatically explicitly included in the whole energy.

It is important that, in this method, all kinds of interactions between points (lattice-lattice, lattice-continuum and continuum-continuum) can be evaluated separately. This is of a particular interest for calculation of the energy of static electron lattices [4]. It is known that the energy of interaction of electron with other electrons, located in the lattice sites, can be estimated by counting these electron pairs, but this energy diverges for infinite lattices. The correct lattice energy could be obtained only after subtracting the interaction energy of an electron with the compensating positive background charge ("jelly"). The same procedure implied for a finite domain, faces some difficulties, as the interaction energy of electron with background charge (calculation of which is simple for infinite lattices) is much more complicated for finite lattices. Applying the total neutrality condition for the spherical ionic lattice domain leads to cancelling the divergent charge misbalance term in the sum of the interaction energy, but not the conditionally convergent sum, associated with the dipole moment of the sphere. We suggest that the domain should be neutral not only globally, but also locally, for every specific geometric sphere inside the domain, which automatically leads to zero dipole moment. The final expression for interaction energy in this case contains only absolutely (i.e. not conditionally) convergent sums. We calculate this energy with a new approach and demonstrate how this energy depends on domain size, which is of a great importance for estimation of the energy dependence on size of small domains, especially for nanocrystals. Despite the fact that there are publications devoted to calculation of Coulomb interaction in finite crystals, often using periodic boundary conditions, but to our knowledge, there is no such analytical calculation for lattice-background interaction energy for finite crystals.

Here (for any analytical central potential) only 1D integral needs to be calculated, after the radial density function is found.

This function depends on geometry of the domain, but not on an interaction potential.

The method presented here allows calculation of different kinds of interaction energy between lattice nodes and between a lattice node and a point in "jelly-like" distributed charge. The core of the method relies on the geometric probability theory (GPT) modified for a finite lattice domain. The GPT traditionally has been used for calculation of the probability density $P(r)$ to find two randomly chosen points to be a distance $r$ apart inside a uniform domain or in a domain with some continuous point's distributions [5-11], but here it was also extended to discrete distributions.

## II. THE INTERNAL ENERGY EQUATION FORMULATION

Let's consider the system of atoms in a finite spherical crystal of radius R interacting by a two-body central potential $U(|\vec{r}_1 - \vec{r}_2|)$. The interaction energy of all *N* atoms of this spatially confined lattice system is generally written as:

$$H = \frac{1}{2}\sum_{i=1}^{N}\sum_{j=1}^{N-1} U(|\vec{r}_i - \vec{r}_j|). \tag{1}$$

In this study we extend the geometric probability techniques [5-11], which is usually applied for the systems with continuous points distribution, for evaluation of this lattice sum (1). The energy Expression (1) in the continuous form may be written as:

$$H = \frac{1}{2}\sum_{i=1}^{N}\sum_{j=1}^{N-1} U(|\vec{r}_i - \vec{r}_j|) = \frac{1}{2}\iint_{V\,V} \rho(\vec{r}_1) U(|\vec{r}_1 - \vec{r}_2|) \rho(\vec{r}_2) d\vec{r}_1 d\vec{r}_2 \tag{2}$$

where $\rho(\vec{r})$ is the microscopic atoms density distribution function [12]:

$$\rho(\vec{r}) = \sum_{\vec{n}\in\inf} \delta\left(\vec{r} - n_1 \cdot \vec{a}_1 - n_2 \cdot \vec{a}_2 - n_3 \cdot \vec{a}_3\right) \tag{3}$$

with $\delta$ being the Dirac delta-function, $\vec{a}_i$ are the basis vectors of the 3-D lattice, $n_i$ - integers, and $V = \frac{4}{3}\pi R^3$ is the total volume of the domain.

## III. GEOMETRIC PROBABILITY APPROACH

According to GPT [5], for any function $U(r)$ its average value taken over a volume of radius $R$ is:

$$\langle U \rangle = \int_0^{2R} U(r) P_r(r) dr \tag{4}$$

where $P_r(s)$ is the probability density to find the distance $r$ between two points inside the domain [5-11]. This function for a uniformly distributed density points inside the sphere [5] is

$$P_r(r) = \frac{\iint \delta(|\vec{x}-\vec{y}|-r) d\vec{x} d\vec{y}}{\iint d\vec{x} d\vec{y}} = \frac{3r_{12}^2}{R^3}\left[1 - \frac{3}{2}\left(\frac{r_{12}}{2R}\right) + \frac{1}{2}\left(\frac{r_{12}}{2R}\right)^3\right] \tag{5}$$

The extension of (5) for non-uniform density distribution was considered by several authors. Here we use one derived by Schleef et al. [6]:

$$\langle U(R) \rangle \equiv \iint d\vec{r}_1 d\vec{r}_2 \rho(\vec{r}_1) U(|\vec{r}_1 - \vec{r}_2|) \rho(\vec{r}_2) =$$
$$= \int_{r_0}^{2R} \left[r_{12}^2 \int d\Omega_{12} \int d^3r_1 \rho(\vec{r}_1) \rho(\vec{r}_{12}+\vec{r}_1)\right] U(|\vec{r}_{12}|) dr_{12} \equiv \tag{6}$$
$$\equiv \frac{1}{P}\int_{r_0}^{2R} G(r_{12};\rho_1,\rho_2) U(|\vec{r}_{12}|) dr_{12},$$

with

$$\vec{r}_{12} \equiv \vec{r}_2 - \vec{r}_1, \quad d^3 r_{12} = d^3 r_2 \tag{7}$$

and

$$P \equiv \int_{r_0}^{2R} G(r_{12}; \rho_1, \rho_2) dr_{12}. \tag{8}$$

is the total number of pairs of nodal points inside the sphere (Fig. 1). The spherical-symmetric density function $\rho(r)$ used in [6], is angle-independent, so $G(r_{12}; \rho_1, \rho_2)$ in (6) transforms into

$$G_R(r_{12}, R) = 16\pi^2 r_{12} \int_{r_{12}/2}^{R} r_1 \rho(r_1) \left\{ \int_{|r_{12}-r_1|}^{r_1} r_2 \rho(r_2) dr_2 \right\} dr_1. \tag{9}$$

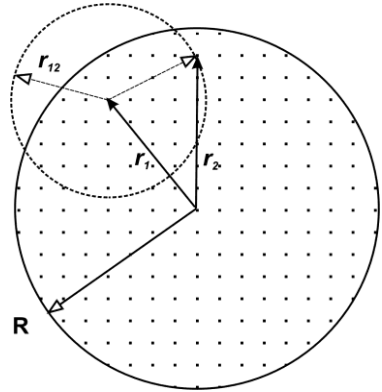

Fig. 1 An example of lattice domain of radius $R$ in the 2-D plane. The circumference of radius $r_{12}$ for a selected point only partly falls inside the domain and therefore number of atoms on this line (which interact with the selected atom inside the domain) differs from such number in the bulk or infinite lattice

In derivation of (9), the symmetry of the integrals, triangle inequality $2r_1 > r_{12}$ (when $r_1 = r_2$) and the law of cosines

$$\cos\Omega_{12} = \frac{r_2^2 - r_{12}^2 - r_1^2}{2r_1 r_{12}} \tag{10}$$

were used. This RDF form keeps all three vectors $\vec{r}_1$, $\vec{r}_2$ and $\vec{r}_{12}$ entirely inside the domain, as evident from Fig. 1. The Expression (6) gives the average energy per one pair of interacting atoms (points), and as there are totally $\frac{N \cdot (N-1)}{2}$ interacting pairs in the domain with $N$ atoms, hence the total internal energy is

$$H = \frac{N \cdot (N-1)}{2} \frac{1}{P} \int_{r_0}^{2R} G(r_{12}; \rho_1, \rho_2) U(|\vec{r}_{12}|) dr_{12}. \tag{11}$$

IV. THE RADIAL DENSITY FUNCTION FOR A SPHERICAL LATTICE DOMAIN

The Formula (9) is applicable only for spherical symmetric density $\rho(r)$, which depends only on the distance $|\vec{r}|$, while the density function $\rho(\vec{r})$ in (2) depends on the whole vector $\vec{r}$. Therefore the approach has to be modified to also account a lattice points distribution. By repeating steps, which lead from Eqs. (6) to (9), but without initial integrations over azimuthal angles $\varphi_1$ and $\varphi_{12}$, the following expression is obtained:

$$G(r_{12}, R) = \frac{2r_{12}^2}{P} \int_{\frac{r_{12}}{2}}^{R} r_1 \int_{\theta_1=0}^{\pi} \left( \int_{\varphi_1=0}^{2\pi} \rho(\vec{r}_1) d\varphi_1 \right) d\theta_1 \left\{ \int_{|r_1-r_{12}|}^{r_1} \left( \int_{\varphi_{12}=0}^{2\pi} \rho(\vec{r}_2) d\varphi_{12} \right) r_2 dr_2 \right\} dr_1. \tag{12}$$

To perform the integrations over the angles in (12), the microscopic density has to be written explicitly in terms of these angles. It might be performed from (3) with a well-known equality:

$$\rho(\vec{r}) = \sum_{\vec{n} \in \inf} \delta(\vec{r} - n_1 \cdot \vec{a}_1 - n_2 \cdot \vec{a}_2 - n_3 \cdot \vec{a}_3) = \frac{1}{v} \sum_{k_1, k_2, k_3 \in \inf} e^{i \cdot 2\pi \cdot (k_1 \vec{b}_1 + k_2 \vec{b}_2 + k_3 \vec{b}_3) \vec{r}} = \frac{1}{v}\left[1 + \sum_{\vec{\gamma} \in \inf}{}' e^{i \cdot 2\pi \cdot \vec{\gamma} \cdot \vec{r}}\right], \quad (13)$$

where $k_1$, $k_2$ and $k_3$ are enumerated by all integer values, $v$ is the volume of the unit cell of the crystal

$$v = \left|\vec{a}_1 \cdot (\vec{a}_2 \times \vec{a}_3)\right| = \left|\vec{a}_2 \cdot (\vec{a}_3 \times \vec{a}_1)\right| = \left|\vec{a}_3 \cdot (\vec{a}_1 \times \vec{a}_2)\right| \quad (14)$$

and $\vec{\gamma} = k_1 \cdot \vec{b}_1 + k_2 \cdot \vec{b}_2 + k_3 \cdot \vec{b}_3$. The symbol $\sum_{\vec{\gamma} \in \inf}{}'$ means that the component with $|\vec{\gamma}| = 0$ is excluded from summation. The basis vectors $\vec{b}_i$ of the reciprocal lattice are expressed via the basis vectors of the real lattice by known relations:

$$\vec{b}_1 = \frac{\vec{a}_2 \times \vec{a}_3}{\vec{a}_1 \cdot (\vec{a}_2 \times \vec{a}_3)}, \quad \vec{b}_2 = \frac{\vec{a}_3 \times \vec{a}_1}{\vec{a}_2 \cdot (\vec{a}_3 \times \vec{a}_1)}, \quad \vec{b}_3 = \frac{\vec{a}_1 \times \vec{a}_2}{\vec{a}_3 \cdot (\vec{a}_1 \times \vec{a}_2)}. \quad (15)$$

In the case of the orthogonal lattice, the axes of the reciprocal lattice are collinear to those of the real crystal lattice [12], simplifying (15) to:

$$b_1 = \frac{1}{a_1}, \quad b_2 = \frac{1}{a_2}, \quad b_3 = \frac{1}{a_3}, \quad (16)$$

and respectively the volume of the unit cell of the crystal to:

$$v = a_1 a_2 a_3, \quad (17)$$

with explicit value of the γ module:

$$|\vec{\gamma}| = \gamma = \left|\sqrt{\left(\frac{k_1}{a_1}\right)^2 + \left(\frac{k_2}{a_2}\right)^2 + \left(\frac{k_3}{a_3}\right)^2}\right|. \quad (18)$$

The sum in (13) could be rewritten using known Rayleigh formula:

$$e^{i 2\pi \vec{\gamma} \vec{r}} = 4\pi \sum_{l=0}^{\infty} \sum_{m=-l}^{l} i^l j_l(2\pi\gamma \cdot r) Y_l^m(\theta', \varphi') Y_l^{*m}(\theta, \varphi) \quad (19)$$

with

$$\gamma_1 = \gamma \cos\varphi' \sin\theta', \quad \gamma_2 = \gamma \sin\varphi' \sin\theta', \quad \gamma_3 = \gamma \cos\theta',$$
$$r_1 = r \cos\varphi \sin\theta, \quad r_2 = r \sin\varphi \sin\theta, \quad r_3 = r \cos\theta, \quad (20)$$
$$\cos\theta' = \frac{k_3}{a_3} \frac{1}{\sqrt{\left(\frac{k_1}{a_1}\right)^2 + \left(\frac{k_2}{a_2}\right)^2 + \left(\frac{k_3}{a_3}\right)^2}}$$

the spherical harmonic function

$$Y_l^m(\theta, \varphi) = \sqrt{\frac{(2l+1)}{4\pi} \frac{(l-m)!}{(l+m)!}} P_l^m(\cos\theta) \cdot e^{im\varphi}, \quad (21)$$

the associated Legendre polynomials $P_l^m(\cos\theta)$ and spherical Bessel functions $j_l(2\pi\gamma r)$. This substitution and transformation lead to the result as:

$$\rho(\vec{r}) = \frac{1}{v}\left[1 + 4\pi \sum_{\vec{k} \in \inf}{}' \sum_{l=0}^{\infty} \sum_{m=-l}^{l} i^l j_l(2\pi\gamma \cdot r) Y_l^m(\theta', \varphi') Y_l^{*m}(\theta, \varphi)\right]. \quad (22)$$

Next, integration of (22) over the azimuthal angle $\varphi$ gives expression:

$$\int_0^{2\pi} \rho(\vec{r})d\varphi = \frac{2\pi}{v} + \frac{1}{v}4\pi \sum_{\vec{\gamma}\in\inf}{}' \sum_{l=0}^{\infty}\sum_{m=-l}^{l} i^l j_l(2\pi\gamma\cdot r)\sqrt{\frac{(2l+1)}{4\pi}\frac{(l-m)!}{(l+m)!}}P_l^m(\cos\theta')\cdot$$

$$\cdot e^{im\varphi'}\sqrt{\frac{(2l+1)}{4\pi}\frac{(l-m)!}{(l+m)!}}P_l^m(\cos\theta)\cdot\int_0^{2\pi} e^{-im\varphi}d\varphi = \qquad (23)$$

$$= \frac{2\pi}{v}\left(1 + \sum_{\vec{\gamma}\in\inf}{}' \sum_{l=0}^{\infty} i^l j_l(2\pi\gamma\cdot r)(2l+1)P_l^0(\cos\theta')\cdot P_l^0(\cos\theta)\right)$$

which is valid for any lattice type. Furthermore, the term with $l=0$ may be separated from the sum in (23) and rearranged using the relation between the associated Legendre polynomials $P_l^m(\cos\theta)$ and the Legendre polynomials

$$P_l^0(\cos\theta) = P_l(\cos\theta). \qquad (24)$$

In this way we receive for (23):

$$\int_0^{2\pi}\rho(\vec{r})d\varphi = 2\pi\rho_s(r) + 2\pi\rho_a(r,\cos\theta). \qquad (25)$$

Here

$$\rho_s(r) = \frac{1}{v}\left[1 + \frac{1}{2\pi r}\sum_{\vec{\gamma}\in\inf}{}' \frac{\sin(2\pi\gamma\cdot r)}{\gamma}\right], \qquad (26)$$

and

$$\rho_a(r,\cos\theta) = \frac{1}{v}\sum_{\vec{\gamma}\in\inf}{}' \sum_{l=1}^{\infty} i^l j_l(2\pi\gamma\cdot r)(2l+1)P_l(\cos\theta')\cdot P_l(\cos\theta), \qquad (27)$$

The term $\rho_u$ is the uniform density distribution on the spherical surface of radius $r$, the isotropic term $\rho_s(r)$ and anisotropic one $\rho_a(r,\cos\theta)$ are the correction for the non-uniform lattice density distribution on the spherical surface which arises from the next terms of $\rho(\vec{r})$ expansion expressed as a sum of spherical waves. With these definitions, the integration over $\theta$ in (12) converges to

$$r^2\int_{\theta=0}^{\pi}\left(\int_{\varphi=0}^{2\pi}\rho(\vec{r})d\varphi\right)\sin\theta d\theta = \frac{4\pi r^2}{v} + \frac{2r}{v}\sum_{\vec{\gamma}\in\inf}{}' \frac{\sin(2\pi\gamma\cdot r)}{\gamma}. \qquad (28)$$

This equation, according to Walfisz-Poisson formula [13], is the expression for the density of the lattice points located on the spherical surface for radius $r$. The total number of points (atoms or ions) inside the spherical domain comes by integrating of (28) by radius coordinate:

$$N = \int_0^R r^2\left[\int_0^{\pi}\left(\int_0^{2\pi}\rho(\vec{r})d\varphi\right)\sin\theta d\theta\right]dr = \frac{4\pi R^3}{3v} - \frac{R}{\pi v}\sum_{\vec{\gamma}\in\inf}{}' \frac{\cos(2\pi\gamma\cdot R)}{\gamma^2} + \frac{1}{2\pi^2 v}\sum_{\vec{k}\in\inf}{}' \frac{\sin(2\pi\gamma\cdot R)}{\gamma^3}. \qquad (29)$$

The Eq. (28) shows the symmetrical part (26) of the density function (22) correctly takes into account the number of the lattice points inside the domain, located on the distance $r$ from any lattice point inside the domain. The number of the lattice points laying on the spherical surfaces only partly located inside the domain (Fig. 1) is determined by the density function (22) which included symmetrical (26) and asymmetrical (27) terms. In the next section we use only symmetrical term approximation for calculation example, i.e. only the first term in serial expansion (22) (with $l=0$). In other words, the number of the lattice points laying which belongs to the intersection of the domain surface and spherical surface of the radius $\vec{r}_{12}$ (Fig. 1) has been counted by averaging of the over last surface. The difference

$$\delta N(R) \equiv N - \frac{4\pi R^3}{3v} = -\frac{R}{\pi v}\sum_{\vec{\gamma}\in\inf}{}' \frac{\cos(2\pi\gamma\cdot R)}{\gamma^2} + \frac{1}{2\pi^2 v}\sum_{\vec{k}\in\inf}{}' \frac{\sin(2\pi\gamma\cdot R)}{\gamma^3} \qquad (30)$$

is caused by the peculiarities of the spherical surface cross-section of the real crystal. This originates from geometric impossibility to fit cubic lattice into any spherical domain with exact number of atoms equivalent to the starting cubic domain, as discussed in [3]. In the case of the crystal formed by positive and negative charges (ionic lattices), this difference causes radical fluctuations of surface charge (even if the infinite crystal has neutral symmetry [3]) and leads to the convergence problems, especially for calculation of the Madelung constant. As J. F. Delord suggested (see note in [3]), summing over spheres works if the missing charge will be added back and the whole sphere will be neutral. It should be noticed that even in the case of the cubic domain the assumption of the uniform positive charge distribution (which is valid for infinite lattice) has to be corrected by considering the different number of the nearest lattice points inside the domain and on the surface. The neutrality of ionic lattice means that non-compensation in electronic charge should be equilibrated by the respective extra ionic charge:

$$\delta N(R) = \delta N_I(R). \tag{31}$$

This is the total neutrality condition for ionic lattices. We also suggest that the sphere is neutral not only in total but also locally, for every sphere with $0 \leq r < R$ radius inside the domain. That leads to the condition that additional charge fluctuations inside must be compensated by the redistributions of positive charges

$$\delta N(\mathrm{r}) - \delta N_I(\mathrm{r}) = 0 \tag{32}$$

and as a consequence

$$\frac{1}{R}\int_0^R \left[\delta N(\mathrm{r}) - \delta N_I(\mathrm{r})\right] dr =$$
$$= -\frac{1}{2\pi^2 v}\sum_{\vec{\gamma}\in\inf}{}'\frac{\sin(2\pi\gamma \cdot R)}{\gamma^3} + \frac{1}{\pi^3 v}\frac{1}{R}\sum_{\vec{\gamma}\in\inf}{}'\frac{\sin(\pi\gamma \cdot R)^2}{\gamma^4} - \frac{1}{R}\int_0^R \delta N_I(\mathrm{r})\, dr = 0, \tag{33}$$

where $\delta N_I(\mathrm{r})$ is the variation in positive charges numbers. This condition means that the sphere has zero dipole moment and zero quadrupole moment as well. It is worth noting that the electrostatic energy of a crystal with non-zero dipole moment (i.e. when the Eq. (32) is not fulfilled) converges only conditionally. We call the Eq. (32) as "local neutrality criterion". Using the designation

$$\delta \tilde{N}(\mathrm{R}) \equiv \frac{1}{R}\int_0^R \delta N(\mathrm{r})\, dr = -\frac{1}{2\pi^2 v}\sum_{\vec{\gamma}\in\inf}{}'\frac{\sin(2\pi\gamma \cdot R)}{\gamma^3} + \frac{1}{\pi^3 v}\frac{1}{R}\sum_{\vec{\gamma}\in\inf}{}'\frac{\sin(\pi\gamma \cdot R)^2}{\gamma^4}. \tag{34}$$

the diverging sum $-\dfrac{R}{\pi v}\sum_{\gamma\in\inf}{}'\dfrac{\cos(2\pi\gamma \cdot R)}{\gamma^2}$ can be written as:

$$-\frac{R}{\pi v}\sum_{\vec{\gamma}\in\inf}{}'\frac{\cos(2\pi\gamma \cdot R)}{\gamma^2} = \delta\tilde{N}(\mathrm{R}) + \delta N(\mathrm{R}) - \frac{1}{\pi^3 v}\frac{1}{R}\sum_{\vec{\gamma}\in\inf}{}'\frac{\sin(\pi\gamma \cdot R)^2}{\gamma^4}. \tag{35}$$

This result correlates with Euler's [14, 15] view on divergent series: "Let us say that the sum of any infinite series is a finite expression from which the series can be derived". Returning to (12) and combining Expressions (12), (25) and (28), we can obtain the RDF expression for a spherical lattice domain:

$$G(r_{12}, R) = \frac{16\pi^2 r_{12}}{P} \int_{r_{12}/2}^{R} r_1 \cdot \rho_s(r_1) \cdot dr_1 \cdot \int_{|r_{12}-r_1|}^{r_1} \left[\rho_s(r_2) + \rho_a(r_2, \frac{r_2^2 - r_{12}^2 - r_1^2}{2r_1 r_{12}})\right] r_2\, dr_2 =$$
$$= G^s(r_{12}, R) + G^a(r_{12}, R), \tag{36}$$

which contains the spherical symmetrical part of the probability density

$$G^s(r_{12}, R) = \frac{16\pi^2 r_{12}}{v^2 P}\int_{r_{12}/2}^{R} r_1 dr_1 \int_{|r_{12}-r_1|}^{r_1} r_2 dr_2 + \frac{16\pi^2 r_{12}}{v^2 P}\int_{r_{12}/2}^{R} r_1 dr_1 \int_{|r_{12}-r_1|}^{r_1} \sum_{\vec{\gamma}\in\inf}{}'\frac{1}{2\pi k}\sin(2\pi k \cdot r_2)\, dr_2 +$$
$$+\frac{16\pi^2 r_{12}}{v^2 P}\int_{r_{12}/2}^{R} \sum_{\vec{\gamma}\in\inf}{}'\frac{1}{2\pi\gamma}\sin(2\pi\gamma \cdot r_1)\, dr_1 \int_{|r_{12}-r_1|}^{r_1} r_2 dr_2 + \tag{37}$$
$$+\frac{16\pi^2 r_{12}}{v^2 P}\int_{r_{12}/2}^{R} \sum_{\vec{\gamma}_1\in\inf}{}'\frac{1}{2\pi\gamma_1}\sin(2\pi\gamma_1 \cdot r_1)\, dr_1 \int_{|r_{12}-r_1|}^{r_1} \sum_{\vec{\gamma}_2\in\inf}{}'\frac{1}{2\pi\gamma_2}\sin(2\pi\gamma_2 \cdot r_2)\, dr_2$$

and asymmetrical one

$$G^a(r_{12}, R) = \frac{16\pi^2 r_{12}}{P} \int_{r_{12}/2}^{R} r_1 \cdot \rho_s(r_1) \cdot dr_1 \cdot \int_{|r_{12}-r_1|}^{r_1} \rho_a(r_2, \frac{r_2^2 - r_{12}^2 - r_1^2}{2 r_1 r_{12}}) r_2 dr_2. \tag{38}$$

The first term in (37) gives for $r_0 = 0$ a well-known result for a uniform distribution of density points ("jelly model" [5], Fig. 2):

$$G_u(r_{12}, R) = \frac{16\pi^2 r^{12}}{v^2 p} \int_{r_{12}/2}^{R} r_1 dr_1 \int_{|r_{12}-r_1|}^{r_1} r_2 dr_2 = \frac{3 r_{12}^2}{R^3}\left[1 - \frac{3}{2}\left(\frac{r_{12}}{2R}\right) + \frac{1}{2}\left(\frac{r_{12}}{2R}\right)^3\right] \tag{39}$$

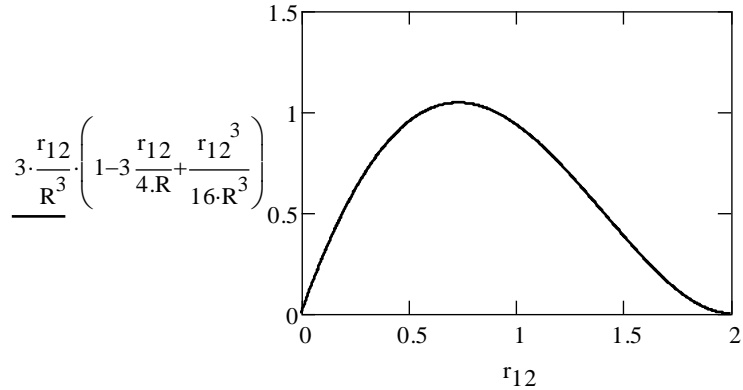

Fig. 2 The plot of the probability density (to find the distance $r$ between two points inside a sphere with uniformly distributed density points as the function of $r_{12}$) for a sphere with unity radius

The subtraction of Eqs. (39) from (36) gives the difference between the total number of pairs points (separated by the distance $r_{12}$) in the lattice and jelly models. Having the expression for $G(r_{12}, R)$ from (36) one can calculate the interaction energy via integration in (11), not only between the lattice points, but also between the lattice points and the jelly points. This technique is demonstrated below by calculation of the interaction energy between electrons, which occupy their lattice sites and compensating positive background charge in simple cubic lattice crystal, limited by a sphere.

V. CALCULATION EXAMPLE: THE INTERACTION ENERGY OF STATIC ELECTRON LATTICES IN SPHERICAL DOMAIN

Here is an example of the calculation of the energy of the static electrons lattice with compensating positive background charge. The interaction energy between the electron and compensating positive background charge from the energy of interaction electrons located in their lattice sites shall be subtracted from the whole energy for proper calculation of the lattice energy (see also the example in [4]). The interaction energy between electron and compensating positive background charge for infinite lattice has a simple form:

$$U_b = \frac{N}{V} \cdot e^2 \lim_{x,y,z \to \infty} \int_0^x \int_0^y \int_0^z \frac{dxdydz}{\sqrt{x^2+y^2+z^2}} \tag{40}$$

owing to translational symmetry of the infinite lattice. In (40) the ratio $\frac{N}{V}$ is the average atomic density $n_d$, and $e$ is the electron charge. This energy diverges in infinite domain [4], and the interaction energy of electrons on their lattice sites

$$U_{ll} = \frac{e^2}{a} \cdot \sum{}' \frac{1}{\sqrt{l^2+m^2+n^2}} \tag{41}$$

also diverges in infinite domain as well [4]. The parameter $a$ is the distance between the lattice sites (the lattice period in a simple cubic lattice). The difference between two energies is, however, finite:

$$U_{ll} - U_b = -2.837297 \frac{e^2}{a} \tag{42}$$

and is approaching the coefficient known as Madelung constant (in this case $M_{SC}$ for simple cubic lattice [4]). The Eq. (40) has such simple form only for infinite lattice, while in a finite lattice the energy depends more strongly on the lattice type and on the position in the domain occupied by the electron, as it is shown in Fig. 1. Therefore, Eq. (40) must be corrected. With this new approach, it could be done directly, using Eqs. (11) and (37):

Our method derived above allows quite simple calculation of such energy for finite lattice domain. This lattice-background Coulomb energy (per particle), comes from the general Expression (11) as

$$U_{fb} = e^2 \frac{N}{2} \int_0^{2R} G_{+-}(r_{12}; \rho_1, \rho_2) \frac{1}{r_{12}} dr_{12}, \tag{43}$$

with $G_{+-}(r_{12}; \rho_1, \rho_2)$ as the RDF for positively-negatively charged pairs. This RDF is the subset of Eq. (36) and we use only symmetrical part of this equation (see the explanation after Eq. (29))

$$G_{+-}(r_{12}, R) = \frac{16\pi^2 r_{12}}{v^2 P} \int_{r_{12}/2}^{R} r_1 dr_1 \int_{|r_{12}-r_1|}^{r_1} r_2 dr_2 + \frac{16\pi^2 r_{12}}{v^2 P} \int_{r_{12}/2}^{R} r_1 dr_1 \int_{|r_{12}-r_1|}^{r_1} \sum_{\vec{\gamma} \in \text{inf}}' \frac{1}{2\pi k} \sin(2\pi k \cdot r_2) dr_2 +$$

$$+ \frac{16\pi^2 r_{12}}{v^2 P} \int_{r_{12}/2}^{R} \sum_{\vec{\gamma} \in \text{inf}}' \frac{1}{2\pi \gamma} \sin(2\pi \gamma \cdot r_1) dr_1 \int_{|r_{12}-r_1|}^{r_1} r_2 dr_2. \tag{44}$$

The first is the RDF for points continuously distributed inside the domain (jelly model). The contribution of this gives the interaction energy between electron and compensating positive background charge for finite lattice sphere (see [6])

$$\frac{N}{V} \cdot e^2 \int_{\text{sphere}} \frac{dxdydz}{\sqrt{x^2+y^2+z^2}} = e^2 \frac{N}{2} \int_0^{2R} \frac{16\pi^2 r_{12}}{v^2 P} \int_{r_{12}/2}^{R} r_1 dr_1 \int_{|r_{12}-r_1|}^{r_1} r_2 dr_2 \frac{1}{r_{12}} dr_{12} = \frac{3}{5} N \frac{e^2}{R}. \tag{45}$$

The second and the third terms in Eq. (44) keep one of the points in every pair exactly in the node of the lattice and there contribution into the energy is

$$e^2 \frac{N}{P} \int_0^{2R} \left\{ \int_{r_{12}/2}^{R} \sum_{\vec{\gamma} \in \text{inf}}' \frac{1}{2\pi\gamma} \sin(2\pi\gamma \cdot r_1) dr_1 \int_{|r_{12}-r_1|}^{r_1} r_2 dr_2 + \int_{r_{12}/2}^{R} r_1 dr_1 \int_{|r_{12}-r_1|}^{r_1} \sum_{\vec{\gamma} \in \text{inf}}' \frac{1}{2\pi\gamma} \sin(2\pi\gamma \cdot r_2) dr_2 \right\} dr_{12}. \tag{46}$$

The last term in Eq. (37) belongs to the lattice-lattice pairs, therefore is not included into Eq. (44). It is worth noting that the lower limit in the integral over $r_{12}$ is zero only for lattice-background pairs, but for lattice-lattice pairs such lower limit is equal to the minimum distance between the lattice nodes. After calculation of integrals in (46) we have

$$-e^2 \frac{N}{P} \left\{ \begin{array}{l} \int_{r_{12}/2}^{R} \sum_{\vec{\gamma} \in \text{inf}}' \frac{1}{2\pi\gamma} \sin(2\pi\gamma \cdot r_1) dr_1 \int_{|r_{12}-r_1|}^{r_1} r_2 dr_2 + \\ + \int_{r_{12}/2}^{R} r_1 dr \int_{|r_{12}-r_1|}^{r_1} \sum_{\vec{\gamma} \in \text{inf}}' \frac{1}{2\pi\gamma} \sin(2\pi\gamma \cdot r_2) dr_2 \end{array} \right\} dr_{12} \simeq$$

$$\simeq -\frac{e^2}{v} \left\{ \sum_{\vec{\gamma} \in \text{inf}}' -\frac{\cos(2\pi\gamma \cdot R)}{\pi\gamma^2} + R^{-2} \sum_{\vec{\gamma} \in \text{inf}}' \frac{3\sin^2(\pi\gamma \cdot R)}{2\pi^3 \gamma^4} - R^{-2} \frac{3}{4\pi^3} \sum_{\vec{\gamma} \in \text{inf}}' \frac{1}{\gamma^4} + R^{-3} \sum_{\vec{\gamma} \in \text{inf}}' \frac{3\sin^2(2\pi\gamma \cdot R)}{8\pi^4 \gamma^5} \right\}. \tag{47}$$

The right-hand side of this equation is the main contribution for the domain size $R \geq 4a$. As it was mentioned before, the sum $\sum_{\vec{\gamma} \in \text{inf}}' -\frac{\cos(2\pi\gamma \cdot R)}{\pi\gamma^2}$ diverges, but by using Eq. (35) we can rewrite Eq. (47) in more suitable form

$$U_{fb} = U_b - \frac{e^2}{v} \frac{1}{2\pi^3} \left\{ R^{-2} \sum_{\vec{\gamma} \in \text{inf}}' \frac{\sin^2(\pi\gamma \cdot R)}{\gamma^4} - R^{-2} \frac{3}{2} \sum_{\vec{\gamma} \in \text{inf}}' \frac{1}{\gamma^4} + R^{-3} \sum_{\vec{\gamma} \in \text{inf}}' \frac{3\sin(2\pi\gamma \cdot R)}{4\pi^3 \gamma^5} \right\} + \frac{\delta \tilde{N}(R) + \delta N(R)}{R}, \tag{48}$$

where the sums in figure brackets converge absolutely, not conditionally. The dipole moment and surface-originating terms $\frac{\delta \tilde{N}(R) + \delta N(R)}{R}$ must be compensated by positive background charge for a correctly defined model. Therefore, the expression

$$\delta U(R) = -\frac{e^2}{v}\left\{\sum_{\vec{\gamma}\in\inf}{}' R^{-2}\sum_{\vec{\gamma}\in\inf}{}' \frac{\sin^2(\pi\gamma\cdot R)}{2\pi^3\gamma^4} - R^{-2}\frac{3}{4\pi^3}\sum_{\vec{\gamma}\in\inf}{}'\frac{1}{\gamma^4} + R^{-3}\sum_{\vec{\gamma}\in\inf}{}'\frac{3\sin(2\pi\gamma\cdot R)}{8\pi^4\gamma^5}\right\} \quad (49)$$

describes explicitly the excess of the interaction energy between electron and compensating positive background charge for finite spherical lattice, in comparison with infinite lattice. As a result of this calculation, Eq. (42) for the Madelung constant for infinite simple cubic lattice transforms into

$$U_{ll} - U_{fb} = -2.837297\frac{e^2}{a} - \delta U(R). \quad (50)$$

Fig. 3 shows the ratio $\frac{\delta U(R)}{M_{SC}}$, which depicts the additional energy contribution to the crystal energy originating from the space limitation. It is seen this extra contribution by order of 0.16-0.002% of the Madelung constant value is in the range $R$ between (5…40) $a$. For particle with $a \sim 0.4$ nm this means $2 < R < 16$ nm. This contribution increases energy of a small particle (i.e. adds extra surface energy), which agrees with well-known tendency of small particles to coarsen by minimizing their energy (decreasing surface). Expression (36) is potential-independent and numerical values are only function of the lattice type and its size.

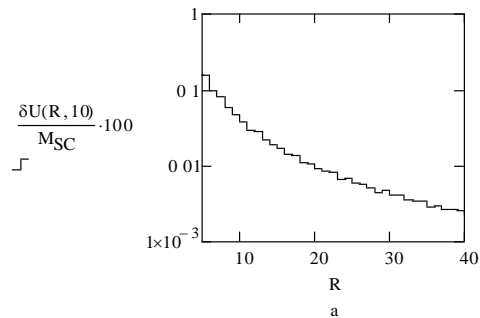

Fig. 3 The plot of the ratio of $\frac{\delta U(R)}{M_{SC}}$ - the excess of the interaction energy between electron and compensating positive background charge to the Madelung constant for finite spherical lattice comparing with infinite lattice as the function of the sphere radius $R$ for the simple cubic crystal in $\frac{R}{a}$ length units

This example shows that extended geometric probability density approach makes it possible correctly calculate interaction between lattice sites in a space-confined systems such as small nanoparticles. It also includes the additional compensating charge, which was previously known causing problems in lattice sums calculations [3].

Our approach can be tailored for calculation lattice energy in non-spherical domain as well, an example for an oblate spheroidal domain. The equation for the surface of an oblate spheroidal domain (an oblate spheroid) is

$$\frac{x^2+y^2}{R_{major}} + \frac{z^2}{R_{minor}} = 1 \quad (51)$$

where $R_{major}$ is the major semi-axe and $R_{minor}$ minor semi-axe. In the spherical coordinates, the equation for an oblate spheroid is [8]

$$R(\theta) = \frac{R_{major}\sqrt{1-\varepsilon^2}}{\sqrt{1-\varepsilon^2\sin^2\theta}} \quad (52)$$

where

$$\varepsilon = \sqrt{1 - \frac{R_{minor}^2}{R_{major}^2}} \quad . \tag{53}$$

The Eq. (6), which is valid for sphere, shall be substituted by the equation

$$\langle U \rangle_\varepsilon \equiv \iint d\vec{r}_1 d\vec{r}_2 \rho(\vec{r}_1) U(|\vec{r}_1 - \vec{r}_2|) \rho(\vec{r}_2) = \int d\Omega_{12} \int_{r_0}^{2R(\theta)} U(|\vec{r}_{12}|) r_{12}^2 dr_{12} \int d^3 r_1 \rho(\vec{r}_1) \rho(\vec{r}_{12} + \vec{r}_1) \tag{54}$$

for an oblate spheroid domain. In this equation the density function $\rho(\vec{r})$ comes from Eq. (22). The Taylor expansion of Eq. (54) in the power of $\varepsilon$ allows simplifying this equation for small $\varepsilon$ (slightly deformed sphere). Similar expression can be written for an arbitrary convex lattice domain, but the function $R(\theta, \varphi)$, which describes the surface of the domain, depends on the two angles $\theta$ and $\varphi$

$$\langle U \rangle_{\{R(\theta,\varphi)\}} \equiv \iint d\vec{r}_1 d\vec{r}_2 \rho(\vec{r}_1) U(|\vec{r}_1 - \vec{r}_2|) \rho(\vec{r}_2) =$$
$$= \int d\Omega_{12} \int_{r_0}^{2R(\theta,\varphi)} U(|\vec{r}_{12}|) r_{12}^2 dr_{12} \int d^3 r_1 \rho(\vec{r}_1) \rho(\vec{r}_{12} + \vec{r}_1). \tag{55}$$

## VI. CONCLUSIONS

In this work, a new method for calculation of the radial density function and geometric probability density is developed for the crystal lattices in a finite spherical shape domain. Extended geometric probability technique allows application of an arbitrary fixed lattice point distributions, leading to explicit expression for the RDF, which is presented as function of the lattice sums. These sums are absolutely converging and are suitable for numeric computations. The internal energy of the system of atoms located at the nodes of a lattice was written as the function of the RDF (36), and it could be applied for any realistic nanoparticles size and interaction potentials. In the example for a simple cubic lattice, the extra contribution to the energy was calculated as an addition to the Madelung constant, representing explicit effect of surface energy and was numerically assessed for different radius of the particle. The excess of the interaction energy between electron and compensating positive background charge for finite spherical lattice, comparing with infinite lattice has been obtained in the form of absolutely convergent lattice sums. The result obtained opens a possibility to evaluate explicitly any realistic lattice type nano-particles, for example, gold nanoparticles, from the point of view of their interaction with environment.

## ACKNOWLEDGMENTS

I would like to thank Prof. Michael Gasik, Aalto University Foundation, for very helpful discussions. Financial support from the Foundation of Helsinki University of Technology and Finnish National Technology and Innovation Agency (Tekes) is gratefully acknowledged.